\documentclass{article}
\usepackage[utf8]{inputenc}
\usepackage{times}
\usepackage{amsfonts}
\usepackage{amsmath,amssymb}
\usepackage{indentfirst}
\usepackage[toc,page]{appendix}
\usepackage{color}
\usepackage{cite}
\usepackage[inactive]{srcltx}
\usepackage[T1]{fontenc}
\usepackage{float}
\usepackage{graphicx}
\usepackage{graphics,color}
\usepackage{amsmath,amssymb}
\usepackage{caption}
\usepackage{subcaption}
\usepackage{wrapfig}
\usepackage{dcolumn}
\usepackage{bm}
\usepackage{hyperref}
\usepackage{epsfig}
\usepackage{latexsym}

\begin{document}
	\date{}
	\begin{center}
		{\Large\bf Quantum entanglement in a four-partite hybrid system containing three macroscopic subsystems}
	\end{center}
	\begin{center}
		{\normalsize C. Corr\^{e}a Jr. and A. Vidiella-Barranco \footnote{vidiella@ifi.unicamp.br}}
	\end{center}
	\begin{center}
		{\normalsize{Gleb Wataghin Institute of Physics - University of Campinas}}\\
		{\normalsize{ 13083-859   Campinas,  SP,  Brazil}}\\
	\end{center}
\begin{abstract}
The combination of different quantum systems may allow the exploration of the distinctive 
features of each system for the investigation of fundamental phenomena as well as for quantum technologies.
In this work we consider a setup consisting of an atomic ensemble enclosed within a laser-driven optomechanical 
cavity, having the moving mirror further (capacitively) coupled to a low frequency LC circuit. This constitutes 
a four-partite optoelectromechanical quantum system containing three macroscopic quantum subsystems of a different 
nature, viz, the atomic ensemble, the massive mirror and the LC circuit. The quantized cavity field plays 
the role of an auxiliary system that allows the coupling of two other quantum subsystems.
We show that for experimentally achievable parameters, it is possible to generate steady state bipartite Gaussian 
entanglement between pairs of macroscopic systems. In particular, we find under which conditions it is possible 
to obtain a reasonable amount of entanglement between the atomic ensemble and the LC circuit, systems that might 
be suitable for constituting a quantum memory and for quantum processing, respectively. For complementarity, 
we discuss the effect of the environmental temperature on quantum entanglement.
\end{abstract}
\section{Introduction}\label{intro}
Recently there have been substantial progresses in the preparation of quantum states of macroscopic systems, and very
often a high degree of control can be reached through their coupling to light. For instance, in an optomechanical 
system, one of the mirrors of an optical cavity is constituted by a massive mechanical oscillator that can be 
coupled to cavity fields via radiation pressure \cite{Milburn2015quantum,Aspelmeyer2014cavity,AspelmeyerRevCavityOptomechanics}. 
Atoms are another type of system whose coupling to the electromagnetic field has been successfully (and precisely)
carried out for some years. In particular, it has been experimentally demonstrated the (collective) coupling of an atomic 
ensemble, i.e. a macroscopic system, to the quantized cavity field \cite{hammerer2009establishing,hammerer2010quantum,Haroche2006Exploring}. 
Another macroscopic quantum system that can be efficiently controlled is the well known quantized LC circuit, basically a superconducting  
electromagnetic oscillator at cryogenic temperatures. Interestingly, such a circuit can be connected to an optical cavity 
if one of the capacitor plates constitutes the moving mirror of the cavity itself (mechanical oscillator) 
\cite{girvin2011superconducting,xiang2013hybrid,blais2020circuit}. 
It is possible to cool those systems down to their fundamental states, \cite{taylor2011laser,malossi2021sympathetic}, 
and characteristic quantum phenomena such as entanglement \cite{genes2008emergence,zhang2017transfer}, a valuable resource for
quantum information processing \cite{horodecki09}, can then arise and be detected.
The macroscopic quantum systems briefly described above admit external controls and can also be coupled to each other, making them suitable 
not only for the development of quantum technologies \cite{xiang2013hybrid,barnett17}, but also for investigating the foundations of 
quantum theory \cite{hofer2016proposal,vivoli2016proposal,lambert2011macrorealism}. We may find in the literature several studies 
related to the macroscopic systems mentioned above, which are in general constituted by three parts. For instance, 
there are discussions involving optomechanical cavities in which the quantized electromagnetic field interacts with an 
atomic ensemble \cite{genes2008emergence}, as well as studies of optomechanical cavities directly coupled to a LC circuit 
\cite{li2020stationary}. Clearly, those advances open up new possibilities for exploring alternative architectures involving 
different types of coupled quantum systems.

In this work, we propose a novel configuration of a composite quantum system, constituted by an atomic ensemble
placed inside an optomechanical cavity. The moving and massive mirror constitutes the plate of a capacitor, i.e., it is itself part of 
an LC circuit. Given that the system as a whole is formed by four distinct, interacting subsystems, we hereafter refer to it as 
an ``hybrid" quantum system \cite{xiang2013hybrid,kurizki2015quantum,han2021microwave,clerk2020hybrid,wallquist2009hybrid}. 
We will mainly focus on the possibilities of bipartite, Gaussian steady-state entanglement involving two of the three macroscopic 
subsystems constituting the proposed arrangement, namely, the atomic ensemble (AE), the mechanical oscillator (MO) and the LC circuit. 
The fourth quantum subsystem, that is, the quantized cavity field, is going to be treated as an ancillary subsystem, that simultaneously 
couples to both the atomic ensemble and the mechanical mirror. Needless to say that a hybrid quantum system of this type is quite flexible 
and one can take advantage of specific features of each subsystem \cite{xiang2013hybrid,kurizki2015quantum,correa2018hybrid}.
In the configuration we are proposing, one could control the optoelectromechanical system independently and in more than one way, namely, 
by manipulating i) a DC bias voltage in the circuit; ii) the cavity detunings and iii) the pump laser power, as we are going to discuss below.
In addition to that, at one end of the array we have the atomic ensemble and at the other end, the LC circuit, the two being 
linked by the field mode and the moving mirror.

Our paper is organized as follows. In Section \ref{model} we introduce the four-partite model with the corresponding Hamiltonian,  
and obtain the steady-state solution for the system via quantum Langevin equations. In Section \ref{results} we present the results for 
the steady state Gaussian entanglement between different bipartitions of the system. We also discuss how we can manipulate the system 
in order to optimize the entanglements. In Section \ref{conclusion} we summarize our conclusions.

\section{Model}\label{model}

\begin{figure}[htb]
  \centering
  \includegraphics[scale=0.45]{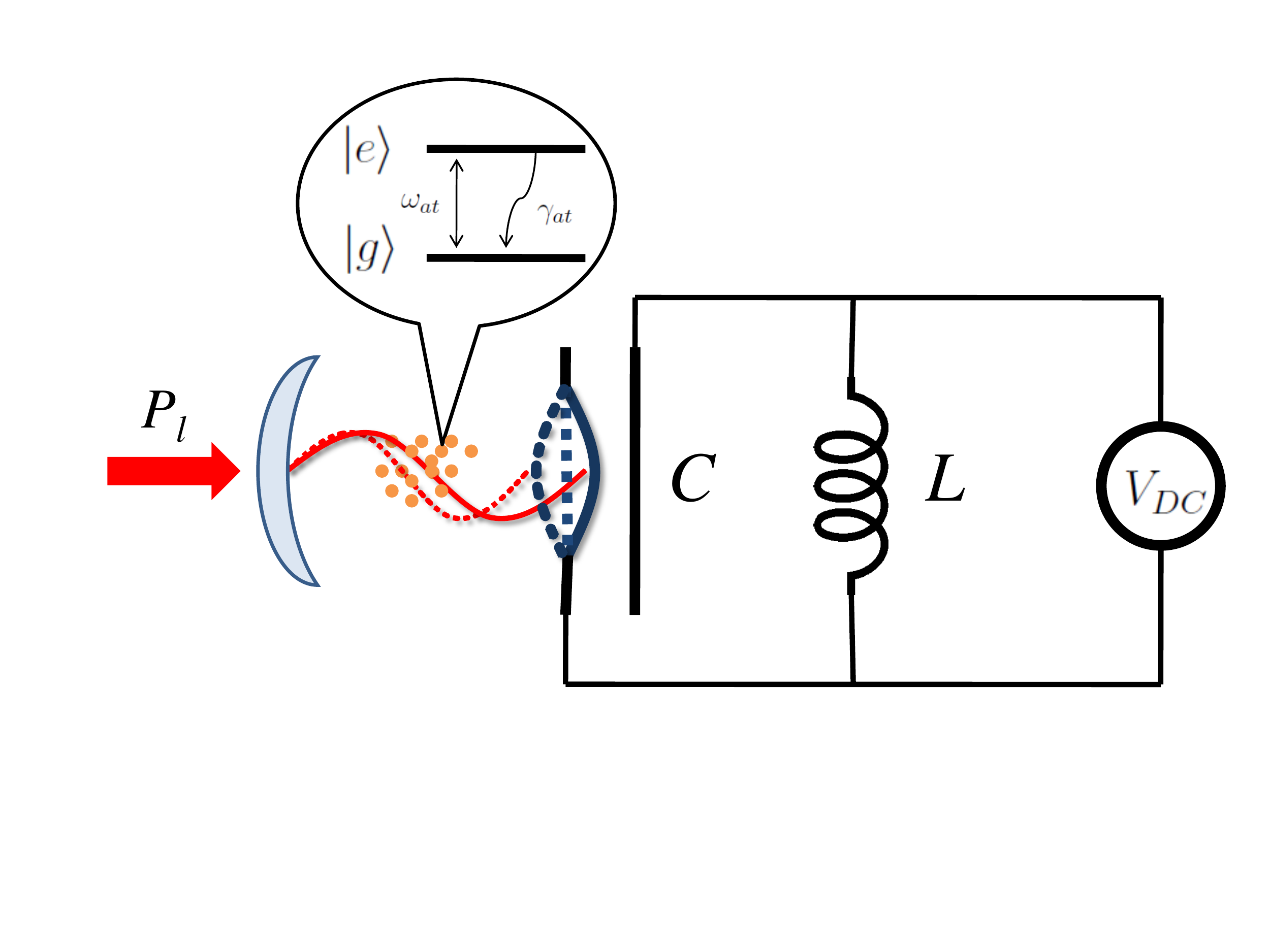}
\caption{Schematic illustration of the proposed setup. An atomic ensemble with $N$ two-level atoms is enclosed within
an optomechanical cavity sustaining a mode of the quantized electromagnetic field of frequency $\omega_{cav}$. 
One of the mirrors can oscillate harmonically with natural frequency $\omega_m$ and also constitutes one of the
plates of a capacitor belonging to an oscillating LC circuit with frequency $\omega_{LC}$. The cavity field is
simultaneously coupled to the atomic ensemble and to the mirror (via radiation pressure), which is 
part of the LC circuit. The cavity is pumped by an external laser of frequency $\omega_l$\label{experimentalsetup} 
and power $P_l$.}
\end{figure}

The system under consideration is constituted by an optomechanical cavity coupled to an LC circuit \cite{li2020stationary,higginbotham2018harnessing}. 
An atomic ensemble containing $N$ two-level atoms is placed inside the cavity \cite{genes2008emergence,murch2008observation}, as shown in Fig. \ref{experimentalsetup}.
Such a four-partite quantum system is described by the following Hamiltonian
\begin{eqnarray} \label{eq:hamiltonian}
	\lefteqn{{\cal{H}}=\hbar\omega_{cav}a^{\dagger}a + \frac{1}{2}\hbar\omega_{at}S_{z} 
	+\frac{1}{2}\hbar\omega_{m}(p^{2}+x^{2}) + \frac{1}{2}\hbar \omega_{LC}(q^{2}+\phi^{2})} \\
&+& \hbar G_{at}(a^{\dagger}S_{-}+aS_{+}) - \hbar G_{om} a^{\dagger}a x + \hbar G_{LC}q^{2}x- q_{0}qV 
+ i\hbar E(a^{\dagger}e^{-i\omega_{l}t}-ae^{i\omega_{l}t}). \nonumber
\end{eqnarray}

The optomechanical cavity of lenght ${\cal L}$ sustains a single mode field of frequency $\omega_{cav}$ with creation (annihilation) 
operators $a^{\dagger}(a)$ (obey $[a,a^{\dagger}]=1$) and has a decay rate $\kappa$. The cavity is also pumped by a laser (classical field) with power $P_{l}$, 
frequency $\omega_l$ and amplitude $E=\sqrt{2P_{l}\kappa/\hbar \omega_{l}}$. The mechanical oscillator (moving mirror), described by the position 
and momentum operators $x$ and $p$ (obey $[x,p]=i\hbar$), has mass $m$, frequency $\omega_m$, a stationary mean position $x^{(s)}$ and zero-point 
fluctuation $x_0 = \sqrt{\hbar/m\omega_{m}}$. The one-photon optomechanical coupling, due to radiation pressure, is given by 
$G_{om}=-\frac{d\omega_{cav}}{dx}|_{x^{(s)}} = (\omega_{cav}x_0/{\cal L})$. The atomic ensemble contains $N$ identical two-level 
atoms with transition frequency $\omega_{at}$, is described by the collective operators $S_{+, -, z}=\sum_{i=1}^{N}\sigma_{+, -, z}^{(i)}$, which 
obey $\left[S_{+}, S_{-}\right]=S_{z}$ and $\left[S_{z}, S_{\pm}\right]=\pm2S_{\pm}$, being $\sigma_{+, -, z}$ the Pauli matrices. 
The atoms are non-resonantly coupled to the cavity field via a collective Tavis-Cummings type of interaction 
\cite{tavis1968exact} with coupling $G_{at}=\mu\sqrt{\omega_{cav}/2\hbar\epsilon_{0}\cal{V}}$. Here $\mu$ is the electric dipole moment of the atomic 
transition, $\epsilon_{0}$ is the electrical permittivity of the vacuum, and $\cal{V}$ the volume of the cavity mode. The quantized LC circuit, 
constituted by a resistance ($R$), a capacitor ($C$) and an inductor ($L$) has a charge operator $q$, a magnetic flux operator $\phi$ (obey $[q,\phi]=i\hbar$) 
and natural resonance frequency $\omega_{LC}=1/\sqrt{LC}$. Here we are going to explore the possibility of having a LC circuit 
with a natural resonance frequency $\omega_{LC}$ close to typical MO frequencies, i.e., in the frequency range of $1$ MHz to $10$ MHz 
\cite{li2020stationary,chu2020perspective}.
One of the capacitor plates is part of the mechanical oscillator itself, and hence its capacitance is a function of position,
$C(x)$. The electromechanical coupling constant is given by $G_{LC}=\frac{q_{0}^{2}q^{(s)}}{\hbar}\frac{d(C^{-1})}{dx}|_{x^{(s)}}$, where 
$q_{0}=\sqrt{\hbar/L\omega_{LC}}$, is the zero-point fluctuation of the capacitor charge and $q^{(s)}$ is the stationary mean charge of the capacitor.
A DC bias voltage $V$ can also be applied to it \cite{bagci2014optical,haghighi2018sensitivity}, allowing an additional control. 
Therefore, we have at our disposal a hybrid quantum system consisting of experimentally accessible macroscopic quantum subsystems that are 
of a different nature, namely, an atomic ensemble, an LC circuit and a mechanical ressonator. Here the quantized cavity field plays the role of an 
auxiliary subsystem that ensures the coupling between the atoms and the mirror.

It is now convenient to apply the Holstein-Primakoff transformation \cite{holstein1940field} to the light-atomic ensemble interaction terms \cite{hammerer2010quantum}, 
which maps fermion operators to boson operators through $ S_{+} = c^{\dagger}\sqrt{N-c^{\dagger}c}\cong \sqrt{N}c^{\dagger} $, 
$ S_{-} = \sqrt{N-c^{\dagger}c}c\cong \sqrt{N}c $ and $ S_{z} = c^{\dagger}c - N/2 $, with $\left[c, c^{\dagger}\right] = 1$. This transformation is valid 
when the average number of atoms in the ground state is much greater than the atoms in the excited state and $N$ is large  
\cite{hammerer2010quantum,genes2008emergence}. Transforming the Hamiltonian Eq.(\ref{eq:hamiltonian}) to the rotating frame of the driving laser 
with frequency $\omega_{l}$ and applying the Holstein-Primakoff transformation, we obtain
\begin{eqnarray} \label{eq:hamiltonianPrime}
	\lefteqn{{\cal{H}'}=\hbar\Delta_{cav}a^{\dagger}a + \hbar\Delta_{at}c^{\dagger}c +\frac{1}{2}\hbar\omega_{m}(p^{2}+x^{2}) 
	+ \frac{1}{2}\hbar \omega_{LC}(q^{2}+\phi^{2})}  \\
&+& \hbar G_{at}'(a^{\dagger}c + ac^{\dagger}) - \hbar G_{om} a^{\dagger}a x + \hbar G_{LC}q^{2}x- q_{0}qV + i\hbar E(a^{\dagger} - a), \nonumber
\end{eqnarray}
where $\Delta_{cav}=\omega_{cav}-\omega_{l}$,  $\Delta_{at}=\omega_{at}-\omega_{l}$ and $G_{at}' = \sqrt{N}G_{at}$. 
We note than an effective coupling $G_{at}' \gtrsim 1$ MHz can be reached with an experimentally reasonable number of atoms, say 
$N \approx 10^{6}$  \cite{kurizki2015quantum}.

In order to study the dynamics of the hybrid optoelectromechanical system plus the atomic ensemble, we may construct a set of 
(nonlinear) quantum Langevin equations in which the dissipation and fluctuation terms are added to the Heisenberg equations of motion arising from 
Eq.(\ref{eq:hamiltonianPrime}), that is

\begin{eqnarray} \label{eq:langevin}
\dot{a} &=& -\left(\kappa + i\Delta_{cav}\right)a - iG_{at}'c + iG_{om}ax + E + \sqrt{2\kappa}a^{in} \nonumber \\
\dot{c} &=& -\left(\gamma_{at} + i\Delta_{at}\right)c - iG_{at}'a + \sqrt{2\gamma_{at}}c^{in} \nonumber \\
\dot{q} &=& \omega_{LC}\phi \nonumber \\
\dot{\phi} &=& -\omega_{LC}q -\gamma_{LC}\phi - 2G_{LC}qx + \frac{q_{0}}{\hbar}V \\
\dot{p} &=& -\omega_{m}x + G_{om}a^{\dagger}a - G_{LC}q^{2} - \gamma_{m}p + \xi \nonumber \\
\dot{x} &=& \omega_{m}p. \nonumber
\end{eqnarray}
Here $\gamma_{at}$, $\gamma_{LC}=2R/L$ and $\gamma_{m}$ are the damping rates for the atoms, the LC circuit and the mechanical 
resonator, respectively. The operator $a^{in}$ ($c^{in}$) describes the input noise affecting the cavity field (atoms) 
\cite{gardiner2004quantum,giovannetti2001phase}, with correlation functions
\begin{equation}
    \left\langle a^{in}(t)a^{in\dagger}(t') \right\rangle = \left\langle c^{in}(t)c^{in\dagger}(t') \right\rangle = \delta(t - t').
\end{equation}
Besides, we assume that the mechanical mode is affected by a stochastic Brownian noise $\xi$. In the limit of a high quality mechanical factor,
$\mathsf{Q}_{m}=\omega_{m}/\gamma_{m} \gg 1$, we can use the Markovian approximation (memoryless correlation) and the Brownian noise $\xi(t)$ 
will obey the relation \cite{Milburn2015quantum,Aspelmeyer2014cavity,gardiner2004quantum}
\begin{equation}
    \left\langle \xi(t)\xi(t') + \xi(t')\xi(t) \right\rangle/2 \cong \gamma_{m}(2\bar{n}_{m} + 1)\delta(t - t'),
\end{equation}
where $\bar{n}_{m}=\left[\exp{\left(\hbar\omega_{m}/k_{B}T\right)} - 1\right]^{-1}$ is the mean thermal excitation number
of the environment at a temperature $T$. 

Therewith the quantum Langevin equations above can be linearized by writing each Heisenberg operator as a sum of its steady-state 
mean value $O^{(s)}$ plus a quantum fluctuation operator with zero-mean value $\delta O$, i.e., $O = O^{(s)} + \delta O$ ($O = a, c, q, \phi, x, p$) \cite{AspelmeyerRevCavityOptomechanics}. This is justified if the cavity is driven by an intense input laser (strong enough power $P_l$), 
so that the steady-state value of both the intra-cavity field and the LC circuit charge become large, i.e., $|a^{(s)}| \gg 1$ 
and $q^{(s)} \gg 1$.
Substituting the expressions above (operator decompositions) into  Eq.(\ref{eq:langevin}), we obtain a set of equations 
for the steady-state values and an additional set of quantum Langevin equations for the fluctuation operators. 
The results for the steady-state mean values of the relevant operators are 

\begin{eqnarray} \label{eq:langevinSteady}
a^{(s)} &=& \frac{E}{\kappa+i\Delta_{cav}' + \frac{(G_{at}')^{2}}{\gamma_{at} + i\Delta_{at}}} \nonumber \\
c^{(s)} &=& -\frac{iG_{at}'a^{(s)}}{\gamma_{at}+i\Delta_{at}} \nonumber \\
q^{(s)} &=& \frac{1}{\omega_{LC}'}\left(\frac{q_{0}\bar{V}}{\hbar} \right) \label{eqn:steadyLangevin} \\
x^{(s)} &=& \frac{1}{\omega_{m}}\left(G_{om}|a^{(s)}|^{2} - G_{LC}(q^{(s)})^{2} \right) \nonumber \\
p^{(s)} &=& \phi^{(s)} =0, \nonumber
\end{eqnarray}
where $\Delta_{cav}'=\Delta_{cav} - G_{om}x^{(s)}$, and $\omega_{LC}'=\omega_{LC} + 2G_{LC}x^{(s)}$ is the new 
resonance frequency of the circuit. We expanded the external bias voltage as the DC bias plus the Johnson-Nyquist 
voltage fluctuation, according to the Fluctuation-Dissipation theorem, i.e,. $V(t)= \bar{V} + \delta V(t)$. For a sufficiently large 
quality factor, or $\mathsf{Q}_{LC}=\omega_{LC}/\gamma_{LC} \gg 1$ we can use the Markov approximation and obtain the autocorrelation 
function $\left\langle \delta V(t)\delta V(t') + \delta V(t')\delta V(t) \right\rangle/2 \cong R\hbar\omega_{LC}(2\bar{n}_{LC} + 1)\delta(t - t')$, 
with $\bar{n}_{LC}=\left[\exp{\left(\hbar\omega_{LC}/k_{B}T_{LC}\right)} - 1\right]^{-1}$ being the mean thermal excitation number in the 
electrical circuit. In general $T_{LC} \ne T $ due to the susceptibility of the LC resonator to electromagnetic and thermal noise 
\cite{malossi2021sympathetic}, but in the remaining of the text we will assume that $T_{LC} \approx T$, given that the (red detuned) cavity field 
cools the mechanical oscillator, which in turn cools the electrical resonator \cite{genes2008robust,genes2008ground}.

In the strong driving condition, we may omit the nonlinear cross fluctuations terms like
$\delta a^{\dagger}\delta a$ and $\delta a\delta x$, obtaining the linearized Langevin equations for the quantum fluctuations operators of the system
\begin{eqnarray} \label{eq:langevinFluctuation}
\delta\dot{a} &=& -\left(\kappa + i\Delta_{cav}'\right)\delta a - iG_{at}'\delta c + iG_{om}a^{(s)}\delta x + \sqrt{2\kappa}a^{in} \nonumber \\
\delta\dot{c} &=& -\left(\gamma_{at} + i\Delta_{at}\right)\delta c - iG_{at}'\delta a + \sqrt{2\gamma_{at}}c^{in} \nonumber \\
\delta\dot{q} &=& \omega_{LC}'\delta\phi \nonumber \\
\delta\dot{\phi} &=& -\omega_{LC}'\delta q -\gamma_{LC}\delta\phi - 2G_{LC}q^{(s)}\delta x + \frac{q_{0}}{\hbar}\delta V \label{eqn:fluctuationLangevin} \\
\delta\dot{p} &=& -\omega_{m}\delta x + G_{om}a^{(s)}(\delta a + \delta a^{\dagger}) - 2G_{LC}q^{(s)}\delta q - \gamma_{m}\delta p + \xi \nonumber \\
\delta\dot{x} &=& \omega_{m}\delta p. \nonumber
\end{eqnarray}

We now define the quadrature fluctuation operators 
$\delta X\equiv (\delta a+ \delta a^{\dagger})/\sqrt{2}$, $\delta Y\equiv (\delta a - \delta a^{\dagger})/i\sqrt{2}$, 
$\delta x_{c}\equiv (\delta c+ \delta c^{\dagger})/\sqrt{2}$, $\delta y_{c}\equiv (\delta c - \delta c^{\dagger})/i\sqrt{2}$, as well as
the input noise operators $X^{in}\equiv (a^{in} + a^{in \dagger})/\sqrt{2}$, $Y^{in}\equiv (a^{in} - a^{in \dagger})/i\sqrt{2}$, 
and $x_{c}^{in}\equiv (c^{in} + c^{in \dagger})/\sqrt{2}$, $y_{c}^{in}\equiv (c^{in} - c^{in \dagger})/i\sqrt{2}$.
This makes  possible to write Eqs.(\ref{eq:langevinFluctuation}) in the following form
\begin{equation}
	\dot{\textbf{u}}(t)=\textbf{A}\textbf{u}(t)+\textbf{n}(t),
\end{equation}
where $\textbf{u}^{T}(t)=(\delta x, \delta p, \delta X, \delta Y, \delta x_{c}, \delta y_{c}, \delta q, \delta \phi)$ is the vector 
associated to the quadrature fluctuations and 
$\textbf{n}^{T}(t)=(0, \xi, \sqrt{2\kappa}X^{in}, \sqrt{2\kappa}Y^{in}, \sqrt{2\gamma}x_{c}^{in}, \sqrt{2\gamma}y_{c}^{in}, 0, q_{0}\delta V/\hbar)$
the vector associated to the input noise operators. $\textbf{A}$ is a $8\times8$  drift matrix which reads
\begin{equation}
\textbf{A} = 
\begin{pmatrix}
0 & \omega_{m} & 0 & 0 & 0 & 0 & 0 & 0 \\
-\omega_{m} & -\gamma_{m} & G_{om}' & 0 & 0 & 0 & -G_{LC}' & 0 \\
0 & 0 & -\kappa & \Delta_{cav}' & 0 & G_{at}' & 0 & 0 \\
G_{om}' & 0 & -\Delta_{cav}' & -\kappa & -G_{at}' & 0 & 0 & 0 \\
0 & 0 & 0 & G_{at}' & -\gamma_{at} & \Delta_{at} & 0 & 0 \\
0 & 0 & -G_{at}' & 0 & -\Delta_{at} & -\gamma_{at} & 0 & 0 \\
0 & 0 & 0 & 0 & 0 & 0 & 0 & \omega_{LC}' \\
-G_{LC}' & 0 & 0 & 0 & 0 & 0 & -\omega_{LC}' & -\gamma_{LC}
\end{pmatrix},
\end{equation}
with $G_{om}'=\sqrt{2}|a^{(s)}| G_{om}$ and $G_{LC}'=2q^{(s)} G_{LC}$.

The linearized Langevin equations in Eq.(\ref{eq:langevinFluctuation}) ensure that when the system is stable, it always reaches a Gaussian state whose entanglement can be completely described by the symmetric covariance matrix $\textbf{V}$, with elements $\textbf{V}_{ij}(t)=\langle \textbf{u}_{i}\textbf{u}_{j} + \textbf{u}_{j}\textbf{u}_{i} \rangle/2 $ for $i,j=1,...,8$. The hybrid system is stable and reaches its steady state only if the real part of all the eigenvalues of 
the drift matrix $\textbf{A}$ are negative. The stability conditions in our system can be found by applying the Routh-Hurwitz criterion \cite{dejesus1987routh,parks1993stability,gradshteyn2014table}.  The resulting expressions are too cumbersome to be shown here, but stability is guaranteed in the forthcoming analysis. 
Once one makes sure that the stability conditions are satisfied, the steady-state correlation matrix can be derived from the following Lyapunov equation

\begin{equation}
	\textbf{A}\textbf{V}+\textbf{V}\textbf{A}^{T}=-\textbf{D},\label{lyapunoveq}
\end{equation}
where $\textbf{D}$ is the diagonal matrix for the damping and leakage rates stemming from the noise correlations, or
\begin{equation}
	\textbf{D}=\text{Diag}\left[0, \gamma_{m}(2\bar{n}_{m}+1), \kappa, \kappa, \gamma_{at}, \gamma_{at}, 0, \gamma_{LC}(2\bar{n}_{LC}+1)\right].
\end{equation}

\section{Results}\label{results}

\begin{figure}[!htb]
\centering
\begin{subfigure}[b]{0.45\textwidth}
\includegraphics[width=\textwidth]{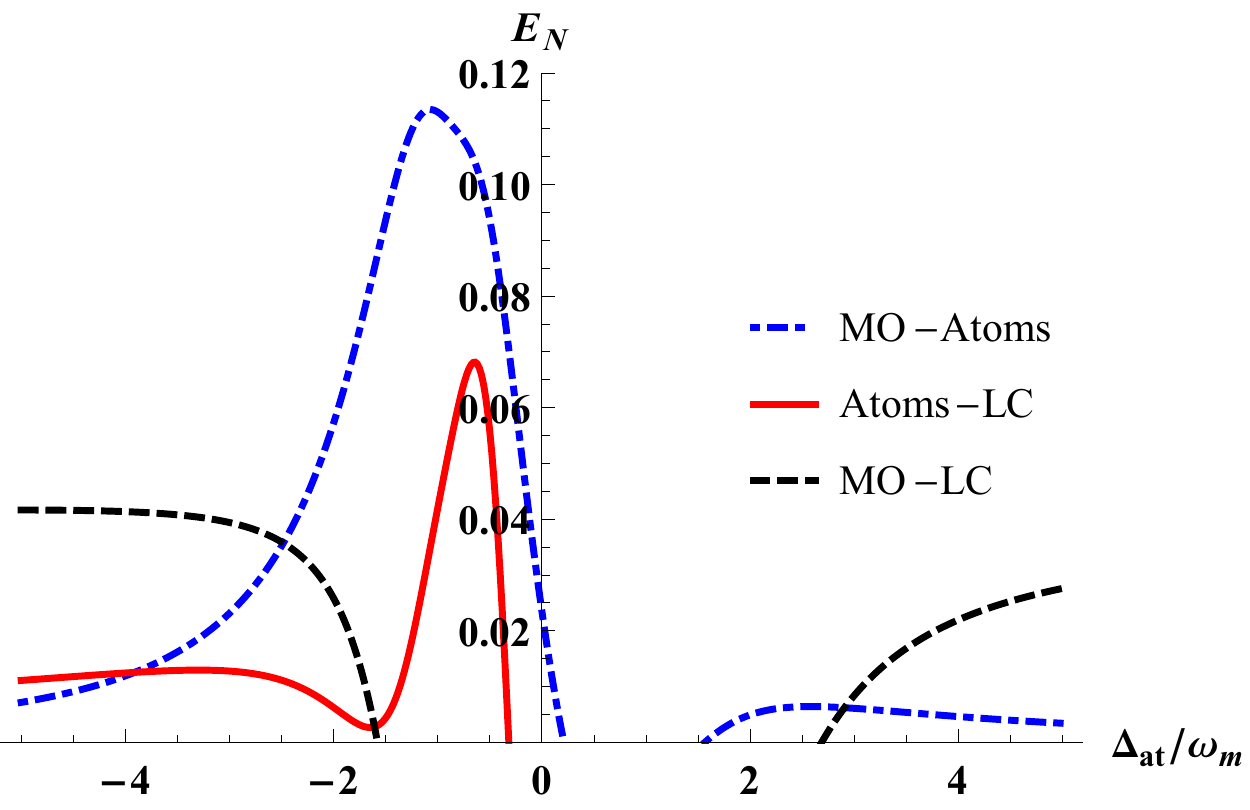}
\caption{}
\end{subfigure}
\hfill
\begin{subfigure}[b]{0.45\textwidth}
\includegraphics[width=\textwidth]{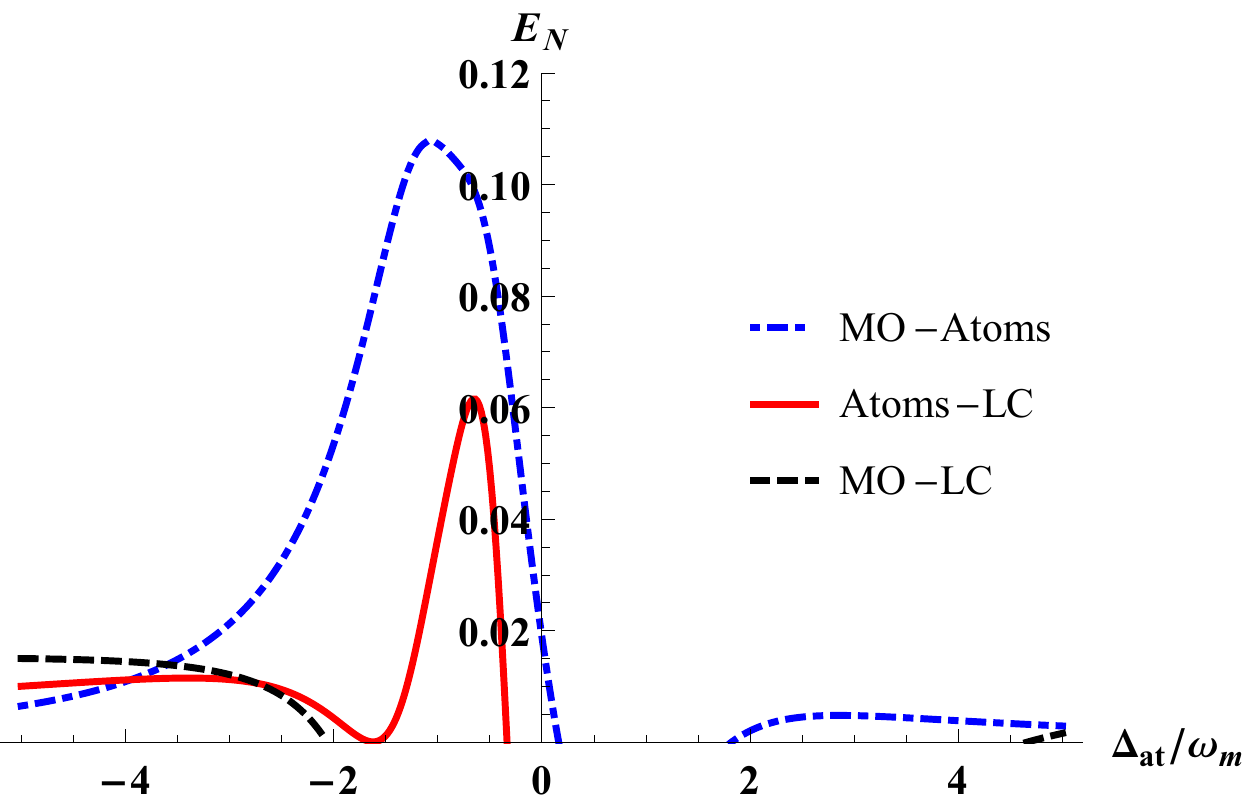}
\caption{}
\end{subfigure}
\caption{Logarithmic negativity relative to the entanglements between: the mechanical oscillator-atomic ensemble 
(blue, dot-dashed curve), the atomic ensemble-LC circuit (red, continuous curve) and the mechanical oscillator-LC circuit 
(black, dashed curve) as functions of the atomic detuning $\Delta_{at}/\omega_{m}$. The environmental temperatures are (a) $T=10$ mK and 
(b) $T=100$ mK. The effective relevant parameters are set as: $\Delta_{cav}' = \omega_{m}$ for the cavity detuning, 
$\omega_{LC}'=\omega_{m}$ for the LC circuit frequency, $G_{LC}'=0.4\omega_{m}$ for LC circuit coupling, and $G_{om}'=0.6\omega_{m}$ 
for the effective optomechanical coupling. The remaining parameters were taken from Table~\ref{tab:experimentalvalues}.}
\label{fig:EN_mac}
\end{figure}

We are primarily interested in the steady-state bipartite entanglement between pairs of subsystems, 
which can be quantified using the logarithmic negativity $(E_{N})$ for two-mode Gaussian states
\cite{adesso2004extremal,braunstein2005quantum,ferraro2005gaussian,serafini2017quantum}
\begin{equation}
	E_{N}=\text{max}\left[0, -\ln2\eta_{-}\right].
\end{equation}
The $8\times8$ covariance matrix $\textbf{V}$ can be reduced to a $4\times4$ submatrix $\textbf{V}_{S}$ of the bipatitions of interest, that is,
\begin{equation}
	\textbf{V}_{S}\equiv
	\begin{pmatrix}
		\textbf{V}_{A}     & \textbf{V}_{C} \\
		\textbf{V}_{C}^{T} & \textbf{V}_{B}
	\end{pmatrix}
\end{equation}
where $\textbf{V}_{A}$, $\textbf{V}_{B}$ and $\textbf{V}_{C}$ are $2\times2$ sub block matrices of $\textbf{V}_{S}$. We obtain the smallest symplectic eigenvalue $\eta_{-} \equiv 2^{-1/2}\left\{\Sigma(\textbf{V}_{S}) - \left[\Sigma(\textbf{V}_{S})^{2}-4\text{det}\textbf{V}_{S}\right]^{1/2}\right\}^{1/2}$, with
$\Sigma(\textbf{V}_{S})\equiv \text{det}\textbf{V}_{A}+\text{det}\textbf{V}_{B}-2\text{det}\textbf{V}_{C}$.
This allows us to evaluate the entanglement between the following macroscopic subsystems: 
i) the mechanical oscillator and the LC circuit (MO-LC),
ii) the mechanical oscillator and the atomic ensemble (MO-AE), and iii) the atomic ensemble and the LC circuit (AE-LC). 
Firstly we numerically compute and plot the logarithmic negativity as a function of the normalized atomic detuning 
$\Delta_{at}/\omega_{m}$ for the various bipartitions, as shown in Fig. \ref{fig:EN_mac}, for bath temperatures (a) $T=10$ mK and (b) $T=100$ mK. 
We have used typical experimental values for the relevant quantities, which are shown in Table \ref{tab:experimentalvalues}. We note 
that the MO-AE entanglement is predominant (blue curve), showing a similar behavior as in Ref.\cite{genes2008emergence}. 
It is also significant for a wide range of atomic detunings and 
has a maximum value for $\Delta_{at} \approx -1.5\omega_{m}$. Interestingly, the second highest peak in the entanglement curve is relative 
to the the AE-LC bipartition, as we note in Fig. \ref{fig:EN_mac} (red curve). On the other hand, the MO-LC entanglement is substantial for 
values of detuning where the MO-AE and LC-AE are small, with the former entanglement decreasing as the latter ones get larger. 
Indeed, the MO-LC entanglement is null in the interval $-1.5\omega_{m} \lesssim \Delta_{at} \lesssim 2.5\omega_{m}$ even for $T=10$ mK. 
In other words, entanglement can also be redistributed among the subsystems simply by changing the atomic detuning. Another noteworthy 
feature of the system is the robustness of both the MO-AE and AE-LC entanglements against thermal noise, as shown in Fig. \ref{fig:EN_mac}(b). 
Nonetheless, the MO-LC entanglement is considerably degraded under the same conditions, due to the fact that the  
oscillation frequencies of both the mechanical oscillator and the LC circuit are relatively low \cite{li2020stationary,malossi2021sympathetic}. 

\begin{figure}[!htb]
\centering
\begin{subfigure}[b]{0.4\textwidth}
\includegraphics[width=\textwidth]{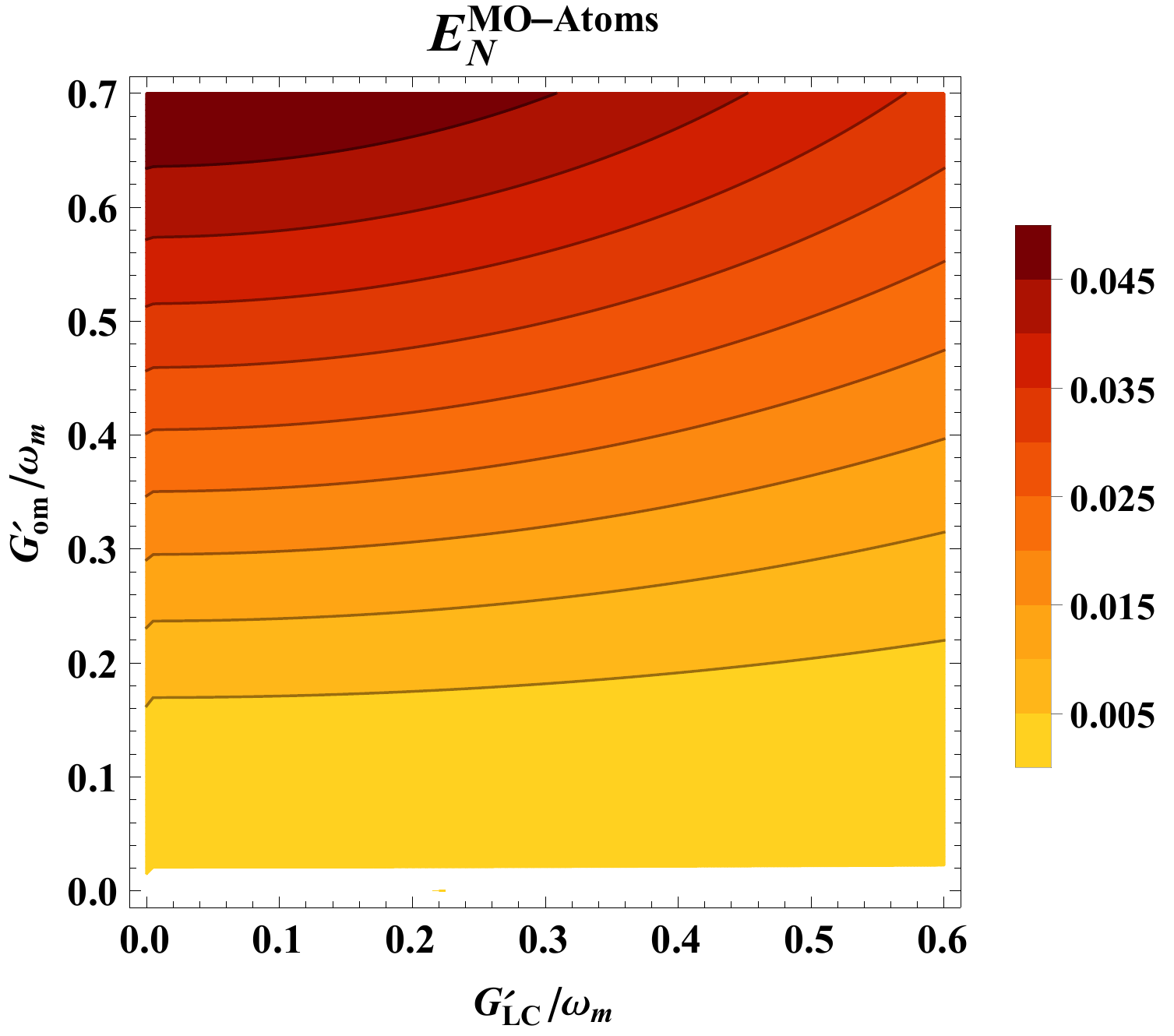}
\caption{}
\end{subfigure}
\hfill
\begin{subfigure}[b]{0.4\textwidth}
\includegraphics[width=\textwidth]{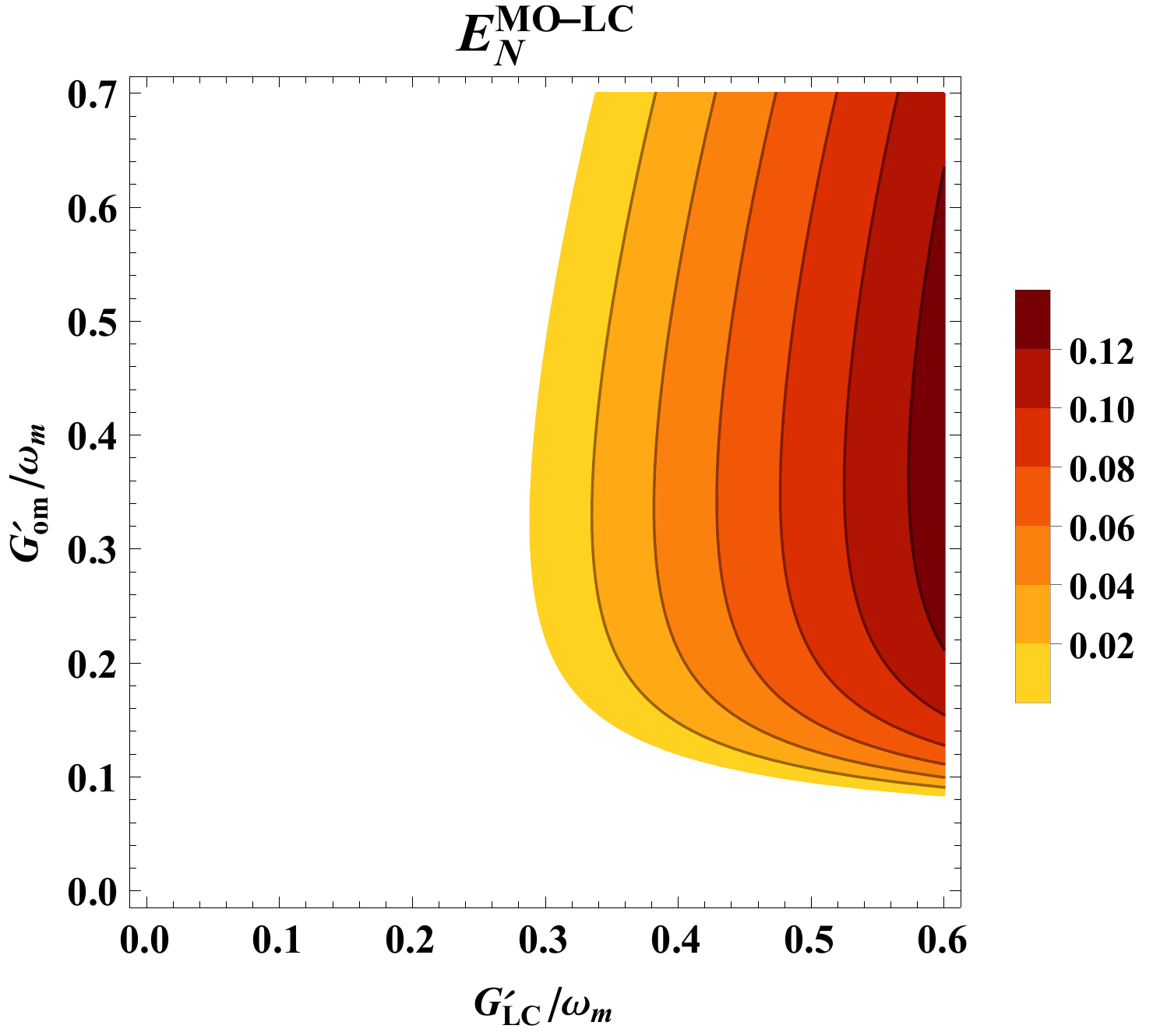}
\caption{}
\end{subfigure}
\hfill
\begin{subfigure}[b]{0.4\textwidth}
\includegraphics[width=\textwidth]{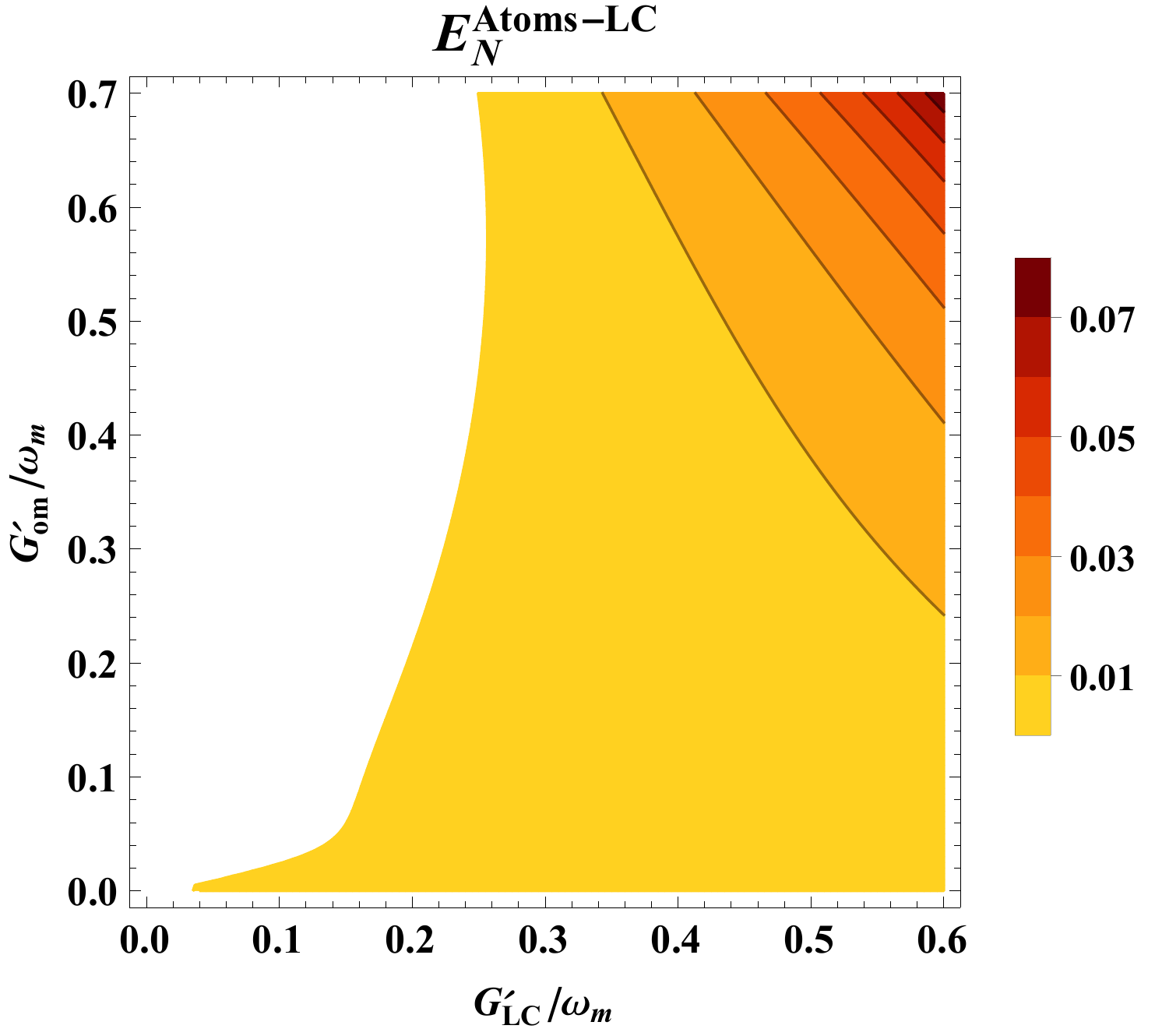}
\caption{}
\end{subfigure}
\hfill
\begin{subfigure}[b]{0.4\textwidth}
\includegraphics[width=\textwidth]{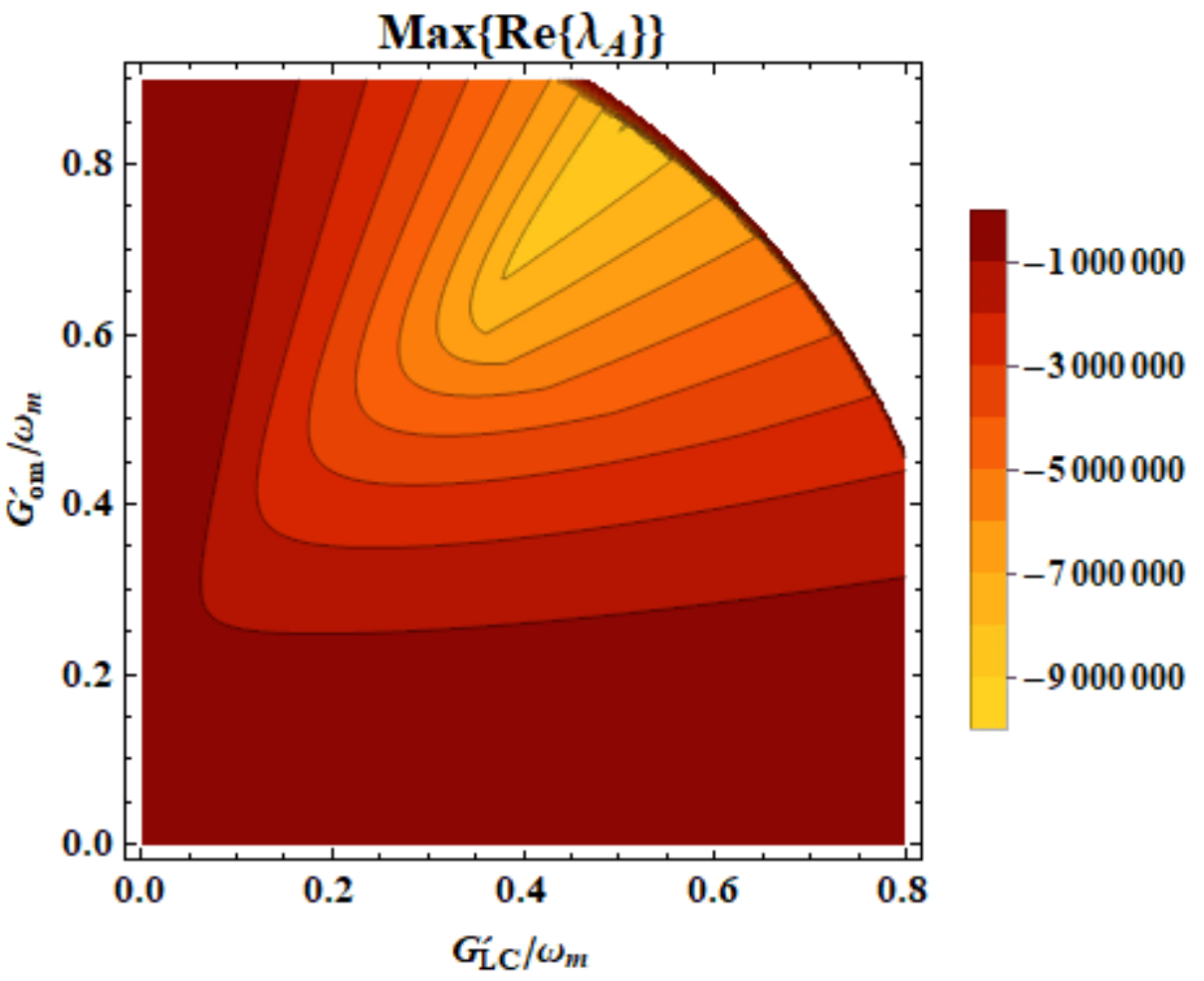}
\caption{}
\end{subfigure}
\caption{Contour plots of the logarithmic negativity relative to the entanglements between: (a) mechanical oscillator-atomic ensemble, 
(b) mechanical oscillator-LC circuit, and (c) atomic ensemble-LC circuit as functions of the effective optomechanical coupling 
$G_{om}'/\omega_{m}$ and the effective LC circuit coupling $G_{LC}'/\omega_{m}$. The white areas in the plots correspond to null entanglement. 
In (d) we have plotted the maximum of the real parts of the eigenvalues of the drift matrix $\textbf{A}$, showing that the stability of the solutions 
is guaranteed within the ranges adopted for the parameters. The white area in the plot shows the instability region. We have 
considered the environmental temperature as being $T=10$ mK, the effective cavity detuning $\Delta_{cav}' = \omega_{m}$, the atomic 
detuning $\Delta_{at}=-2.5\omega_{m}$, and the effective LC circuit frequency $\omega_{LC}'=\omega_{m}$. The remaining parameters were taken from Table~\ref{tab:experimentalvalues}.}
\label{fig:EN_mac_map}
\end{figure}

The degree of entanglement between subsystems should also depend on the effective coupling constants. This is clearly shown in 
Fig. \ref{fig:EN_mac_map}, where we have plotted the logarithmic negativity $E_N$ relative to the following bipartitions: 
(a) MO-AE; (b) MO-LC; and (c) AE-LC, as a function of the normalized effective coupling strengths, $G_{om}'/\omega_{m}$ and $G_{LC}'/\omega_{m}$
and atomic detuning, $\Delta_{at}=-2.5\omega_{m}$. We have picked this specific value of detuning because
all three bipartitions exhibit entanglement for $\Delta_{at}=-2.5\omega_{m}$. 
It is evident in Fig. \ref{fig:EN_mac_map}(a) that the entanglement between the mechanical oscillator and the atomic ensemble has
a strong dependence on the effective optomechanical coupling, that is, $E_N$ gets larger as $G_{om}'$ increases, 
as one would expect. On the other hand $G_{LC}'$ should be kept small enough to favour the entanglement between the 
atomic ensemble and the mirror. In Fig. \ref{fig:EN_mac_map}(b), the entanglement between the mechanical oscillator and the LC 
circuit gets larger as the coupling strength $G_{LC}'$ is increased, although is less sensitive to changes in $G_{om}'$. 
For sufficiently high values of both $G_{LC}'$ and $G_{om}'$ (in very narrow ranges), as shown in Fig. \ref{fig:EN_mac_map}(c), 
it is possible to obtain a significant amount of entanglement between the atomic ensemble and the LC circuit, the two subsystems 
placed in opposite ends in the setup. The effective couplings, and as a consequence,
the entanglement in the system can be tuned by changing experimentally controlled parameters, viz., $G_{om}'$ depends on
the laser pump amplitude $E$, and $G_{LC}'$ has dependence on the DC bias voltage $V$.

Our aim in Fig. \ref{fig:EN_mac_map} was to show the amounts of entanglements that could be simultaneously reached
in the system. Despite the fact that there is no set of parameters that maximizes the entanglement in all three  bipartitions at the 
same time, as we see in Fig. \ref{fig:EN_mac}, it is possible to obtain larger values of entanglement for every case separately if we 
choose appropriate values of couplings and detunings. Our simulations show that there can be reached the following maximum values of 
entanglement (logarithmic negativity): $E_{N} \approx 0.17$ relative to the mechanical oscillator and atomic ensemble (MO-AE) for 
$0.6\omega_{m} \lesssim G_{om}' \lesssim 0.7\omega_{m}$, $G_{LC}' \lesssim 0.2\omega_{m}$, $G_{at}= 0.6\omega_{m}$ and $\Delta_{at}=-\omega_{m}$; 
$E_{N} \approx 0.14$ relative to the mechanical oscillator and the LC circuit (MO-LC) for 
$0.2\omega_{m} \lesssim G_{om}' \lesssim 0.7\omega_{m}$, $0.5\omega_{m} \lesssim G_{LC}' \lesssim 0.6\omega_{m}$, 
$0.2\omega_{m} \lesssim G_{at} \lesssim 0.6\omega_{m}$ and $-3\omega_{m} \lesssim \Delta_{at} \lesssim -2\omega_{m}$. Lastly,  
$E_{N} \approx 0.22$ relative to the atomic ensemble and the LC circuit (AE-LC) for $0.6\omega_{m} \lesssim G_{om}' \lesssim 0.7\omega_{m}$, 
$0.5\omega_{m} \lesssim G_{LC}' \lesssim 0.6\omega_{m}$, $0.4\omega_{m} \lesssim G_{at} \lesssim 0.6\omega_{m}$ and $\Delta_{at} = -\omega_{m}$.
All the other parameters are the same as in Fig. \ref{fig:EN_mac_map}

\begin{figure}[!htb]
\centering
\begin{subfigure}[b]{0.45\textwidth}
\includegraphics[width=\textwidth]{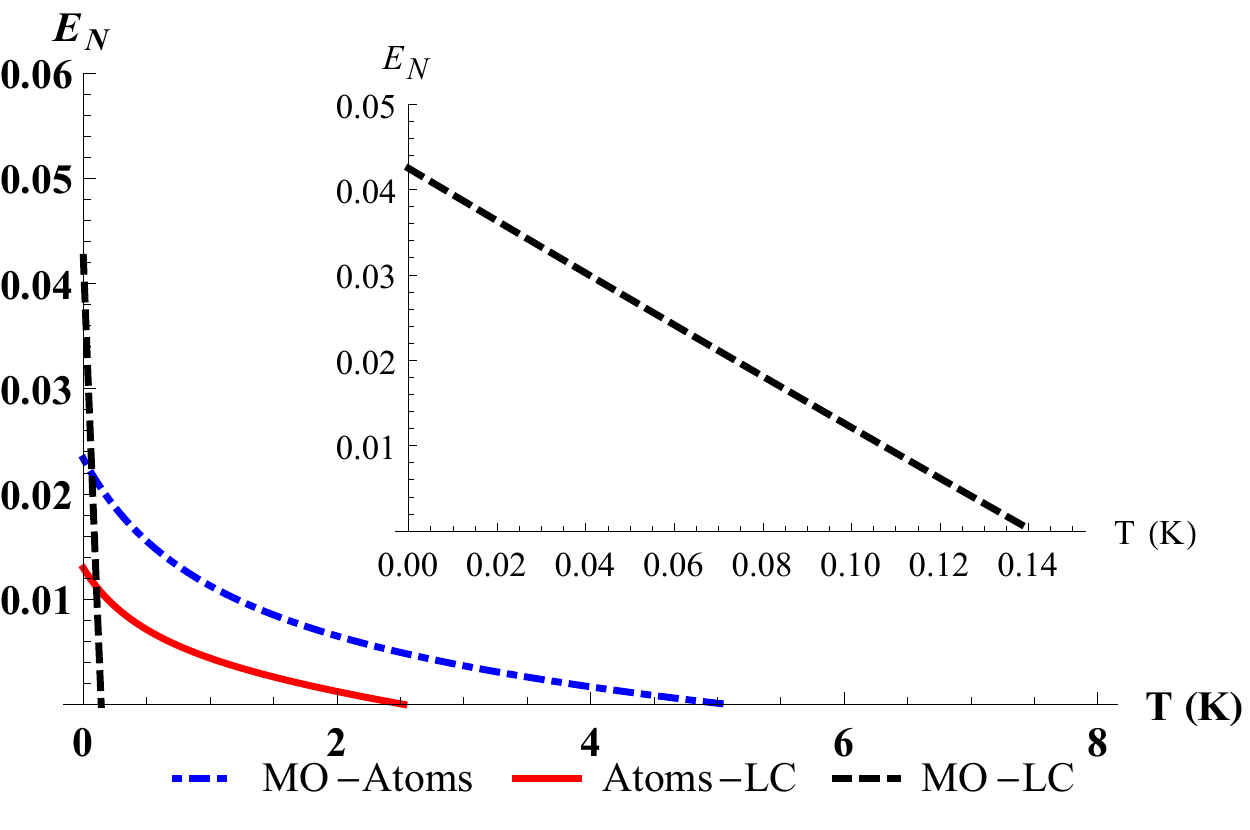}
\caption{}
\end{subfigure}
\hfill
\begin{subfigure}[b]{0.45\textwidth}
\includegraphics[width=\textwidth]{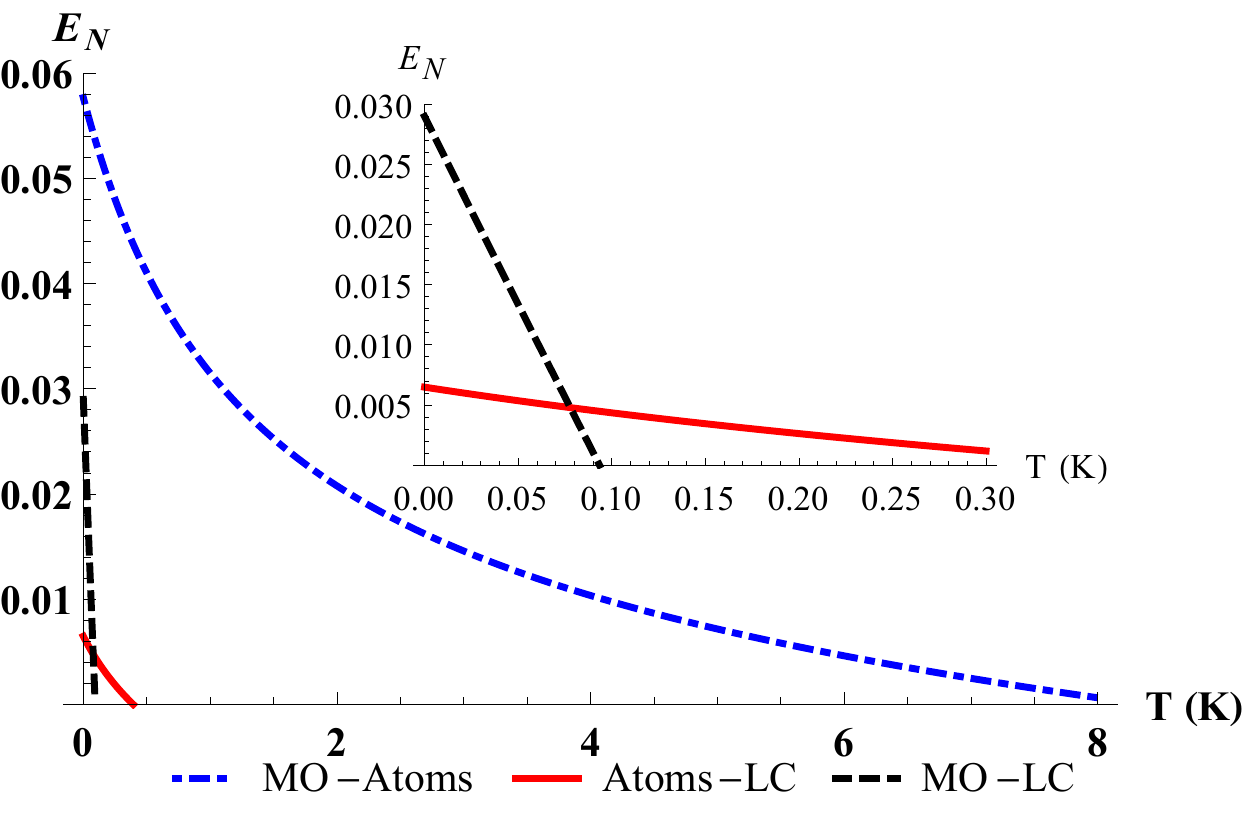}
\caption{}
\end{subfigure}
\caption{Logarithmic negativity relative to the entanglements between: the mechanical oscillator-atomic ensemble (blue, dot-dashed curve), atomic ensemble-LC circuit 
(red, continuous curve) and mechanical oscillator-LC circuit (black, dashed curve) as functions of environmental temperature $T (K)$ for atomic detunings: 
(a) $\Delta_{at}=-3\omega_{m}$ and (b) $\Delta_{at}=-2\omega_{m}$. We have set the effective parameters as: the LC circuit coupling $G_{LC}'=0.4\omega_{m}$, 
the  optomechanical coupling $G_{om}'=0.6\omega_{m}$, and the cavity detuning $\Delta_{cav}' = \omega_{m}$. The remaining parameters were taken from Table~\ref{tab:experimentalvalues}.}
\label{fig:EN_mac_temp}
\end{figure}

\begin{figure}[!htb]
\centering
\begin{subfigure}[b]{0.4\textwidth}
\includegraphics[width=\textwidth]{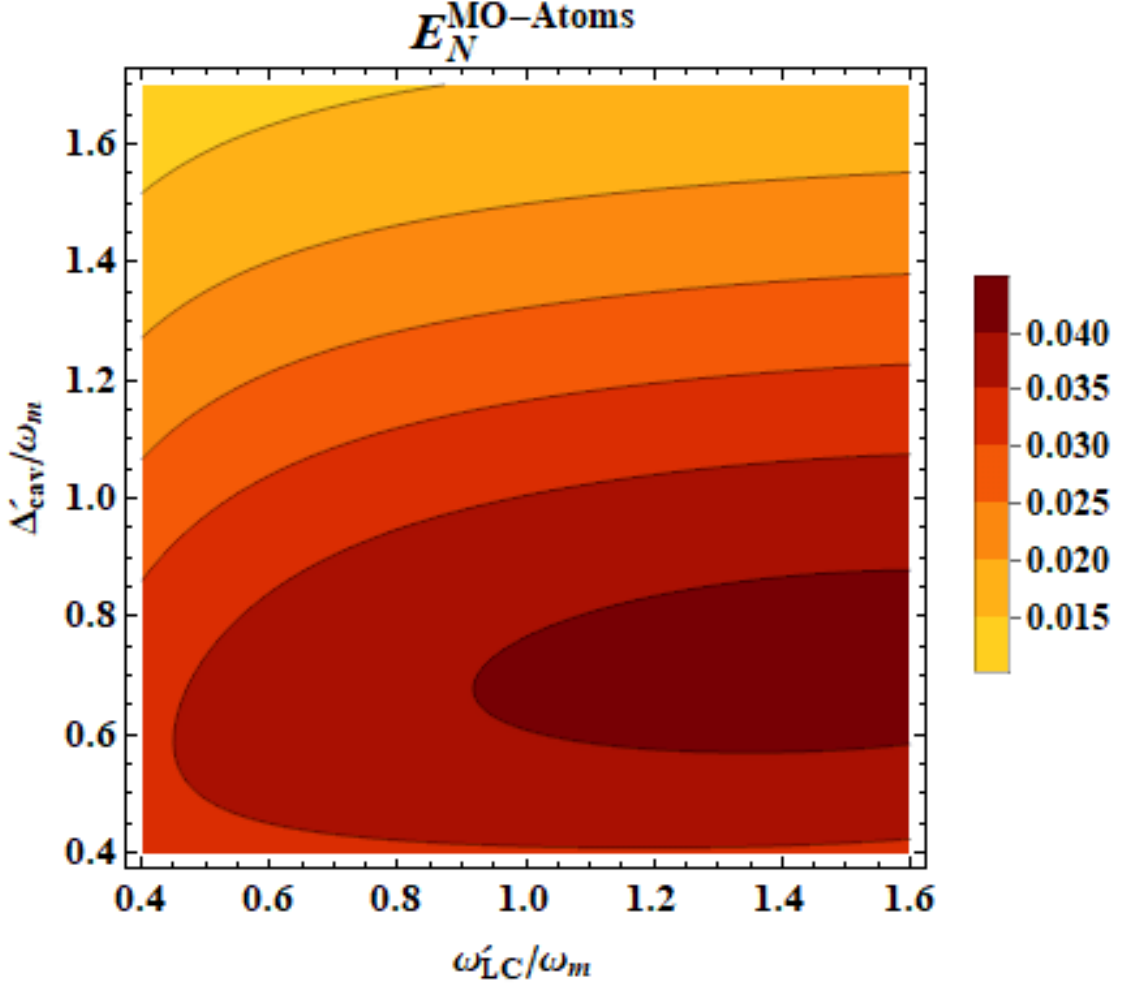}
\caption{}
\end{subfigure}
\hfill
\begin{subfigure}[b]{0.4\textwidth}
\includegraphics[width=\textwidth]{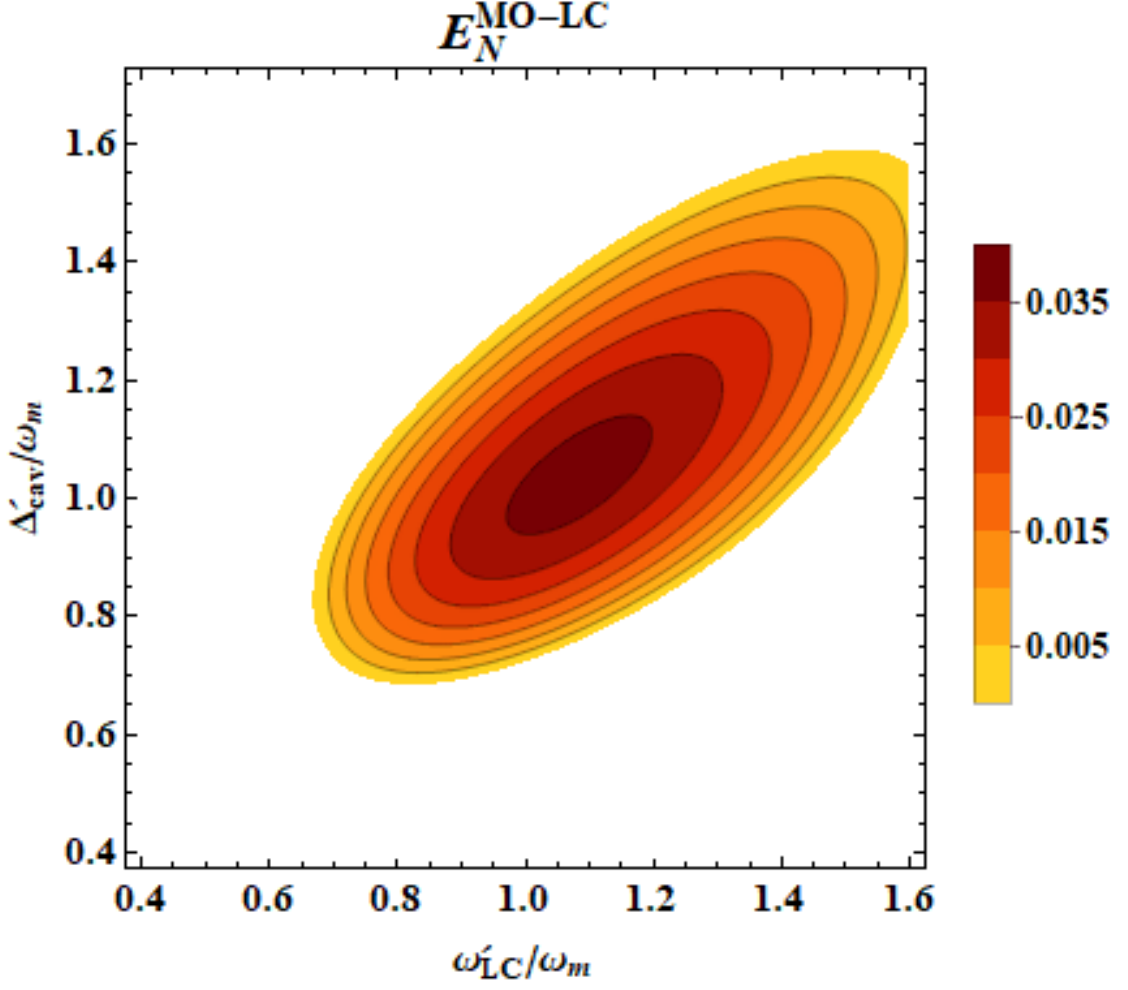}
\caption{}
\end{subfigure}
\hfill
\begin{subfigure}[b]{0.4\textwidth}
\includegraphics[width=\textwidth]{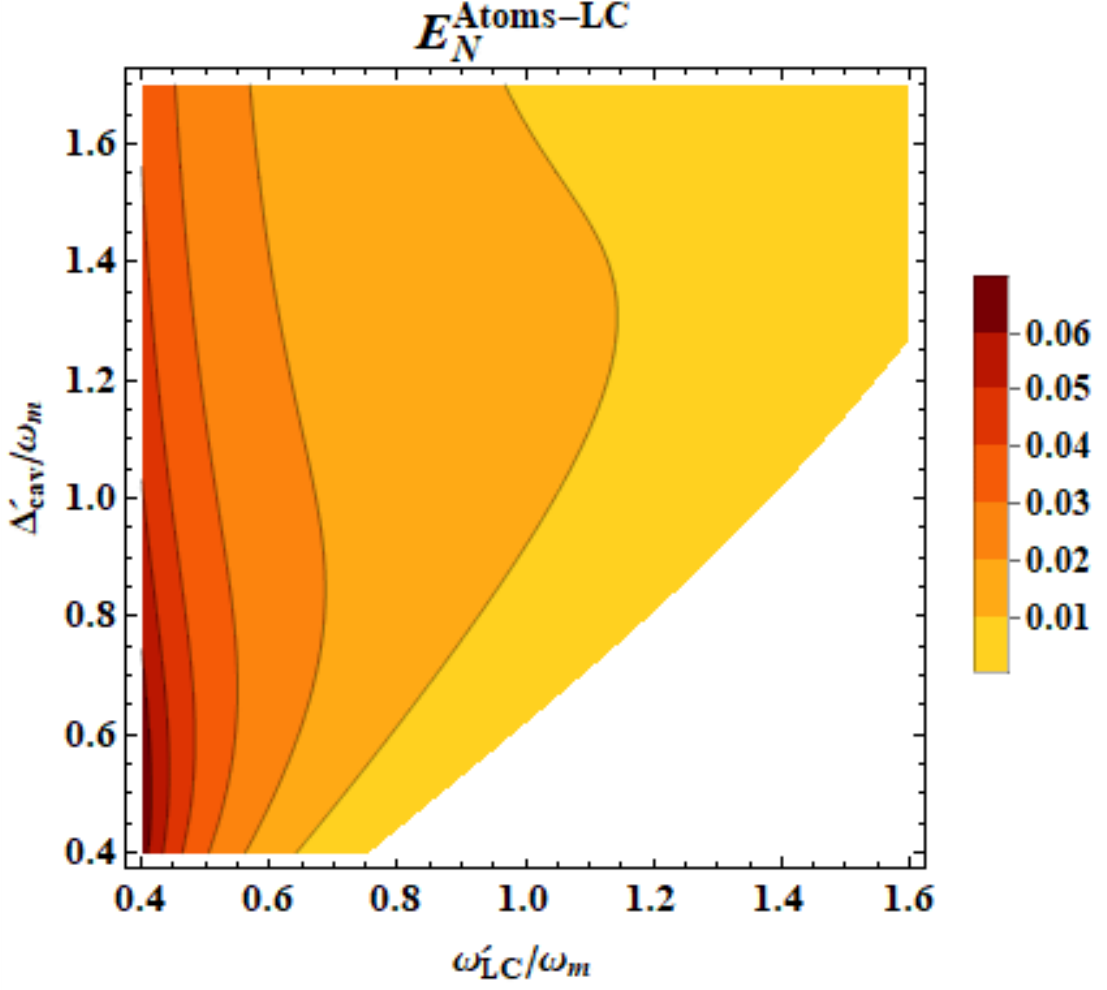}
\caption{}
\end{subfigure}
\hfill
\begin{subfigure}[b]{0.4\textwidth}
\includegraphics[width=\textwidth]{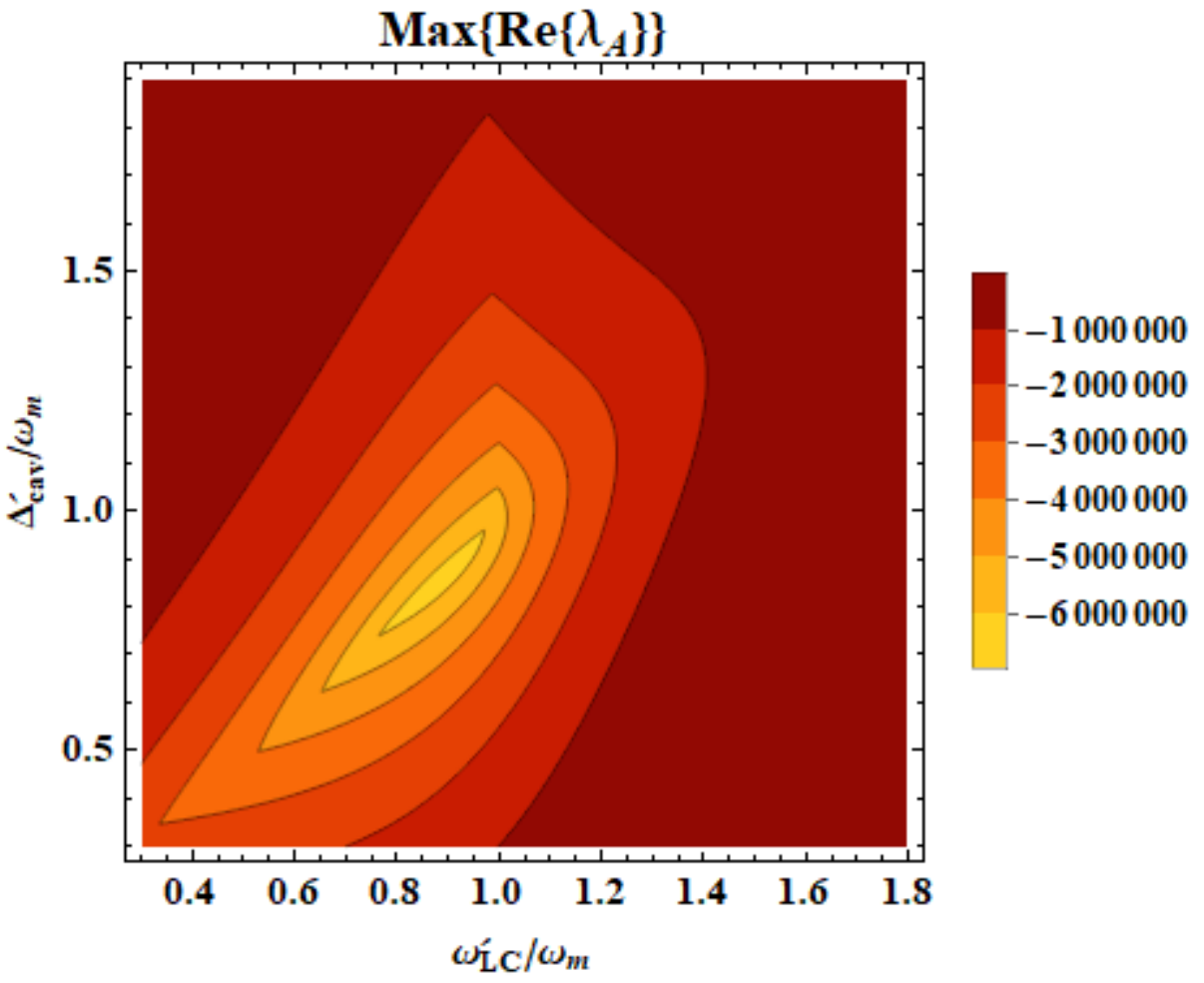}
\caption{}
\end{subfigure}
\caption{Contour plots of the logarithmic negativity relative to the entanglements between: (a) mechanical oscillator-atomic ensemble, 
(b) mechanical oscillator-LC circuit, and (c) atomic ensemble-LC circuit as functions of the effective cavity-laser detuning 
$\Delta_{cav}'/\omega_{m}$ and the effective LC circuit frequency $\omega_{LC}'/\omega_{m}$. The white areas in the 
plots correspond to null entanglement. In (d) we have plotted the maximum of the real parts of the eigenvalues of the drift matrix 
$\textbf{A}$, showing that the stability of the solutions is guaranteed within the ranges adopted for the parameters. We have also 
considered the environmental temperature as being $T=10$ mK, the atomic detuning $\Delta_{at}=-2.5\omega_{m}$ and the effective coupling 
constants being $G_{LC}'=0.4\omega_{m}$ (LC circuit coupling) and $G_{om}'=0.6\omega_{m}$ (optomechanical coupling).
The remaining parameters were taken from Table~\ref{tab:experimentalvalues}.}
\label{fig:EN_mac_lowT}
\end{figure}

There are different degrees of robustness against thermal noise, depending on the subsystems we consider
and the values of the parameters in play. In Fig. \ref{fig:EN_mac_temp} we have plots of the logarithmic negativity relative to the: i) MO-AE; 
ii) AE-LC and iii) MO-LC subsystems, as a function of the environmental temperature and for two different values of the atomic detuning $\Delta_{at}$. 
For $\Delta_{at}=-3\omega_{m}$ we notice that the MO-LC entanglement is the largest one for very low temperatures, but drops off abruptly as the 
temperature is increased, while both the MO-AE and AE-LC entanglements are more robust against thermal noise. Yet, if $\Delta_{at}=-2\omega_{m}$, 
the entanglements between the MO-LC and AE-LC are smaller and go to zero at low temperatures, while the MO-AE entanglement, apart from being the 
largest, is also less susceptible to thermal noise \cite{genes2008emergence}. We may now have a glance at the dependence of the 
entanglement between different subsystems depending on the effective frequencies and detunings. In Fig. \ref{fig:EN_mac_lowT} we display plots of 
the logarithmic negativity, as a function of $\omega'_{LC}/\omega_m$ and $\Delta'_{cav}/\omega_m$, for the three bipartitions here considered. 
We note that the MO-LC circuit entanglement is maximum for near-resonant systems, i.e., 
$\omega'_{LC} \approx \omega_m\,(\Delta'_{cav} \approx \omega_m)$ as shown in Fig. \ref{fig:EN_mac_lowT}(b),
a similar behavior as reported in Ref.\cite{li2020stationary}. The situation
might be different for the other partitions, that is, with respect to the MO-AE and AE-LC entanglement, as shown in Figs. \ref{fig:EN_mac_lowT}(a)
and \ref{fig:EN_mac_lowT}(c). The field-mediated, maximum values of MO-AE entanglement still occur for a relatively wide range of values of 
frequency $\omega'_{LC} \geq \omega_m$ and $0.6\omega_m \lesssim \Delta'_{cav} \lesssim 0.8\omega_m$. A comparable amount of (maximum) 
entanglement can also be generated between the AE-LC subsystems, for $\omega'_{LC} \approx 0.4\omega_m$. We should point out 
that the maximum and minimum values (range) of the parameters being varied in all plots, e.g., the coupling constants, frequencies and 
detunings are such that the solutions found (steady states) are stable, as illustrated in Fig. \ref{fig:EN_mac_map}(d) and Fig. \ref{fig:EN_mac_lowT}(d), 
where we have plotted the maximum of the real parts of the eigenvalues of the drift matrix $\textbf{A}$.

The quantum entanglement generated in the hybrid system as described above is also a potential resource for quantum technologies 
\cite{kurizki2015quantum,wallquist2009hybrid,rabl2006hybrid}. The proposed architecture combines a quantum memory having a long coherence time 
(atomic ensemble) with a fast processor (LC circuit). As shown in  Fig. \ref{fig:EN_mac}(a) we may control the AE-LC circuit entanglement via the 
atomic detuning $\Delta_{at}$. For instance, for $\Delta_{at}\approx -0.6\omega_{m}$ the AE-LC circuit entanglement is maximum, while for 
$\Delta_{at} > -0.3\omega_{m}$ the corresponding quantum states become separable. Thus, the information processed in the LC circuit 
\cite{xiang2013hybrid} could, in principle, be stored (retrieved) in (from) the atomic ensemble \cite{rabl2006hybrid} simply by changing the 
atomic detuning.

\begin{table}
	\centering
		\begin{tabular} {|c|c|c|c|}
		\hline
		\textbf{Parameters}                  &    & \textbf{Values}      \\ \hline
		Cavity length             & ${\cal L}$                & 1 mm            \\ \hline
		Cavity decay              & $\kappa$           & $\pi\times10^{7}$ Hz  \\ \hline
		Pump laser wavelength      & $\lambda$    & 1064 nm          \\ \hline
		Laser power               & $P_l$              & 35 mW \\ \hline
		Mechanical frequency                & $\omega_{m}$         & $2\pi\times10^{7}$ Hz \\ \hline
		Mirror mass             & $m$                 &    10 ng              \\ \hline
		Mechanical damping rate & $\gamma_{m}$        & $200\pi$ Hz       \\ \hline
		Atomic damping rate        & $\gamma_{at}$        & $\pi\times10^{7}$ Hz  \\ \hline
		LC circuit damping         & $\gamma_{LC}$        & $200\pi$ Hz  \\ \hline
		Atomic ensemble-field coupling     & $G_{at}'$              & $1.2\pi\times10^{7}$ Hz  \\ \hline
				\end{tabular}
	\caption{Parameters used in the calculations \cite{genes2008emergence,li2020stationary}.}
	\label{tab:experimentalvalues}
\end{table}

\section{Conclusion}\label{conclusion}

We have demonstrated the feasibility of the generation of steady state entanglement in a quantum system composed of three distinct macroscopic 
subsystems, namely, an atomic ensemble, an optomechanical cavity (massive mirror) and a (low frequency) LC circuit, 
for realistic parameters (see Table \ref{tab:experimentalvalues}). The moving mirror is assumed to be directly coupled to the circuit, while the 
(single mode) electromagnetic quantized field is simultaneously coupled to both the mirror and the atomic ensemble. We usually 
find in the literature proposals for entangling two macroscopic quantum systems, viz.: i) atomic ensemble and massive mirror \cite{genes2008emergence}; 
ii) LC circuit and massive mirror \cite{li2020stationary}, but here we discussed for the first time bipartite entanglement in a system comprising three 
macroscopic subsystems of different type in the same setup. Coupling multiple macroscopic quantum systems certainly presents a challenge, but the
control of pairs of such subsystems previously achieved can facilitate their adequate integration. Each one of the quantum 
subsystems considered here has its peculiarities, which can be advantageous (or not) depending on the 
intended applications. We have shown under which conditions it is possible to generate entangled states between the mechanical oscillator and 
the atomic ensemble; between the  mechanical oscillator and the LC circuit, and remarkably, between the atomic ensemble and the LC 
circuit. The latter case is particularly interesting, that is, the establishment of entanglement between an AE, suitable 
for a quantum memory, and the LC circuit. Both subsystems admit independent external controls and the degree of entanglement
between them can be changed by adjusting the atomic detuning $\Delta_{at}$ or/and the DC bias voltage $V$.

\section*{Acknowledgements}
The authors would like to thank CNPq (Conselho Nacional para o 
Desenvolvimento Cient\'\i fico e Tecnol\'ogico), Brazil,
for financial support through the National Institute for Science and 
Technology of Quantum Information (INCT-IQ under grant 465469/2014-0).

\bibliographystyle{unsrt}
\bibliography{BibliografiaBibTex}

\newpage

\end{document}